\documentclass[sn-mathphys-num]{sn-jnl}% Math and Physical Sciences Numbered Reference Style 
\usepackage{graphicx}%
\usepackage{multirow}%
\usepackage{amsmath,amssymb,amsfonts}%
\usepackage{amsthm}%
\usepackage{mathrsfs}%
\usepackage[title]{appendix}%
\usepackage{xcolor}%
\usepackage{textcomp}%
\usepackage{manyfoot}%
\usepackage{booktabs}%
\usepackage{algorithm}%
\usepackage{algorithmicx}%
\usepackage{algpseudocode}%
\usepackage{listings}%
\usepackage{bm}
\def\tri{{{}^3{\rm H}}}
\def\het{{{}^3{\rm He}}}
\def\heq{{{}^4{\rm He}}}

\def\bmr{{\bm r}}

\def\bmp{{\bm p}}
\def\bmk{{\bm k}}
\def\bmq{{\bm q}}

\def\bmP{{\bm P}}

\def\np{\phantom{0}}

\def\jac{\xi}
\newcommand{\jacb}{{\bm \xi}}
\def\hypfi{\varphi}

\begin{document}

\title{Study of the alpha-particle monopole transition form factor%\thanks{Grants or other notes
}


\author*[1]{\fnm{M.}\sur{Viviani}}\email{michele.viviani@pi.infn.it}
\author[1]{\fnm{A.}\sur{Kievsky}}\email{alejandro.kievsky@pi.infn.it}
\author[2,1]{\fnm{L.E.}\sur{Marcucci}}\email{laura.elisa.marcucci@unipi.it}
\author[3,4]{\fnm{L.}\sur{Girlanda }}\email{girlanda@le.infn.it}


\affil*[1]{\orgdiv{Sezione di Pisa}, \orgname{Istituto Nazionale di Fisica Nucleare},
           \orgaddress{\street{Largo B. Pontecorvo 3}, \city{Pisa}, \postcode{I-56127}, \country{Italy}}}

\affil[2]{\orgdiv{Department of Physics ``E. Fermi''}, \orgname{University of Pisa},
           \orgaddress{\street{Largo B. Pontecorvo 3}, \city{Pisa}, \postcode{I-56127}, \country{Italy}}}

\affil[3]{\orgdiv{Department of Mathematics and Physics}, \orgname{University of Salento},
           \orgaddress{\street{Via Arnesano}, \city{Lecce}, \postcode{I-73100}, \country{Italy}}}

\affil[4]{\orgdiv{Sezione di Lecce}, \orgname{Istituto Nazionale di Fisica Nucleare},
           \orgaddress{\street{Via Arnesano}, \city{Lecce}, \postcode{I-73100}, \country{Italy}}}


\abstract
  {
  The $\heq$ monopole form factor is studied  by computing the transition matrix element
  of the electromagnetic charge operator between the
  $\heq$ ground-state and the $p+\tri$ and $n+\het$ scattering states.  The nuclear wave functions
  are calculated using the hyperspherical harmonic method, by starting from
  Hamiltonians including two- and three-body forces derived in chiral effective field theory.
  The electromagnetic charge operator retains, beyond the leading order (impulse approximation) term, also higher order contributions, as relativistic corrections 
  and meson-exchange currents.
  The results for the monopole form factor are in fairly agreement with recent MAMI data.
  Comparison with other theoretical calculations are also provided.
  }

\maketitle

\section{Introduction}
\label{intro}
The $\heq$ nucleus is a fundamental system for our comprehension of nuclear forces.
The four nucleons form a ground state of quantum numbers $J^\pi=0^+$, hereafter denoted as the $0^+_0$ state.
Such a state is rather deeply bound, with a binding energy of about $7$ MeV per nucleon.  In addition, the $\heq$ nucleus has also
a number of excited states, which, however, are not true bound states but resonances.
The first excited state $0^+_1$ (which has the same quantum numbers as the ground state) is, in fact,
unstable for the splitting in the $p+\tri$ subsystems. It lies approximately 20 MeV above the ground state,
but below the opening of the $n+\het$ channel~\cite{A4b}. Clearly, for the description of this resonance,
the Coulomb interaction plays a very important role~\cite{Gatto24}.

The nature of such a resonance is still a puzzle after many years of studies~\cite{E23}. The process
$\heq(e,e')X$ can be used to obtain  direct information on the monopole form factor $F_M(q)$
(which is essentially the matrix element of the electromagnetic transition operator between the initial and final states),
detecting scattered electrons which have lost approximately $20$ MeV of energy.
The experiments performed in the past~\cite{W70,Fea65,Kea83} could not achieve
a great accuracy. However, more recently, an experiment performed at the Mainz Microtrom (MAMI)
allowed to extract quite accurate data for $F_M(q)$~\cite{Kegel23}.
In this experiment, electron beams with energies of $450$,
$690$, and $795$ MeV were directed onto a target consisting of cryogenic helium gas.
The scattered electrons were detected using a sophisticated apparatus,
which allowed to observe both the elastic peak and the first-excited state resonance.
The $\heq$ elastic peak was used to determine quite accurately the 
luminosity and to estimate the experimental resolution needed for the precise
extraction of the monopole form factor (however, a careful analysis of the data is
necessary in order to subtract the non-resonant contributions, see Ref.~\cite{Kegel23} for more details). 

In the past years, several theoretical studies of
$F_M(q)$ were also performed.
In Ref.~\cite{Hiyama04}, $F_M(q)$ was calculated using a
bound state technique, expanding the wave functions over a 
Gaussian basis. The result of this calculation is in good agreement with
both the old and the MAMI experimental data. In Refs.~\cite{Bacca13,Bacca14,Kegel23}, a calculation
using the Lorentz Integral Transform (LIT) method to sum implicitly all the
intermediate states was performed. In this case, the calculated monopole form factor was
found to be rather at variance with respect to the experimental data, in particular with respect to
the precise MAMI data~\cite{Kegel23}. More recently, a calculation performed using 
the no-core Gamow shell model method, including explicit $p+\tri$, $n+\het$, and $d+d$ reaction channels, allows to reproduce the
MAMI data~\cite{Michel23}. A similar conclusion was found in another recent calculation performed in the framework of
nuclear lattice effective field theory~\cite{M23}.

These calculations were performed using different Hamiltonians. 
In Ref.~\cite{Hiyama04}, the results were obtained 
using the Argonne V8 (AV8')~\cite{AV18} nucleon-nucleon (NN)
potential plus a simple three-nucleon (3N) interaction.
The calculations of Refs.~\cite{Bacca13,Bacca14,Kegel23} were performed
using a NN+3N interaction derived within the framework of the chiral effective field theory ($\chi$EFT).
The Hamiltonian used in the calculation of Ref.~\cite{Michel23}
is based on the $V_{low-k}$ version of the same NN interaction used in Refs.~\cite{Bacca13,Bacca14,Kegel23},
adopting the cutoff value $\Lambda=1.9$ fm${}^{-1}$~\cite{Bogner03}, but without including any 3N force. 
Finally, in the calculation performed in the framework of nuclear lattice effective field theory~\cite{M23},
a rather simple nuclear interaction has been used, in practice containing
NN+3N contact terms. Anyway, this interaction is
capable of reproducing the ground state properties of light nuclei, medium-mass nuclei,
and neutron matter simultaneously with no more than a few percent error in the energies
and charge radii~\cite{M23}. 

In order to analyse further this process, we have reconsidered the study of the monopole transition form factor, 
exploiting our expertise in the calculation of four-nucleon (4N) scattering wave functions. Using modern realistic Hamiltonians we have computed the transition matrix element between the
$\heq$ ground-state and the $p+\tri$ and $n+\het$ scattering states including in the transition operator terms beyond the leading order (impulse approximation), as relativistic corrections and meson-exchange contributions. This calculation follows somewhat the experimental technique itself, where the electrons scatters the $\heq$ nuclei, producing final states composed by different clusters as $p+\tri$, $n+\het$, etc. To compute $F_M(q)$ we have
to integrate over the possible final states of different energies produced in the process.

We have performed this study using the NN chiral interaction derived at next-to-next-to-next-to-leading order (N3LO)
by Entem and Machleidt~\cite{EM03,ME11}, with cutoffs $\Lambda=500$, $600$ MeV. We have also performed calculations including
the chiral 3N interaction derived at next-to-next-to leading order (N2LO) in Refs.~\cite{Eea02,N07}. The two free parameters
in this N2LO 3N potential, denoted usually as  $c_D$ and $c_E$, have been fixed in order to reproduce the experimental values of the $A=3$ binding energies
and the Gamow-Teller matrix element of the tritium $\beta$ decay~\cite{GP06,GQN09}.
Note that these parameters have been recently redetermined~\cite{Mea12,Bea18,Mea18}.

The 4N wave functions are calculated  using the so-called hyperspherical harmonics (HH) technique~\cite{rep08,fip19}.
Several benchmarks performed in the past have shown the good
accuracy which can be achieved by this method for both the 4N bound state problem~\cite{Kea01,Viv05}
and the 4N scattering problem~\cite{bm11,bm17}. The detailed
application of the HH method to the 4N scattering problem is reported in Ref.~\cite{Viv20},
where the convergence issues are throughout discussed and several results for
$n+\tri$, $p+\het$, $p+\tri$, and $n+\het$ scattering are given. 

This paper is organized as follows. In Section~\ref{sec:theo}, a 
description of the HH method for bound and scattering states is briefly resumed,
and the approach for calculating $F_M(q)$ is presented. The results are given in Section~\ref{sec:res}, 
where the comparison with the experimental data and the results of other theoretical calculations is
reported. The conclusions and the perspectives of this
work will be given in Section~\ref{sec:conc}.

\section{Theoretical formalism}
\label{sec:theo}
This section is divided into three  subsections. 
In the first two, we briefly introduce the HH method to compute 4N bound and scattering states. In the last subsection we present the approach employed to calculate $F_M(q)$. 

\subsection{The HH method}
\label{sec:hh}

We start with the definition of the Jacobi vectors which,
for a system of four identical particles (disregarding the proton-neutron mass difference), are given by
\begin{eqnarray}
   \jacb_{1p}& = & \sqrt{\frac{3}{2}} 
    \left ({\bm r}_l - \frac{ {\bm r}_i+{\bm r}_j +{\bm r}_k}{3} \right )\ , \nonumber\\
   \jacb_{2p} & = & \sqrt{\frac{4}{3}}
    \left ({\bm r}_k-  \frac{ {\bm r}_i+{\bm r}_j}{2} \right )\ , \label{eq:JcbV}\\
   \jacb_{3p} & =& {\bm r}_j-{\bm r}_i\ , \nonumber
\end{eqnarray}
where $p$ specifies a permutation corresponding to the order $i$, $j$,
$k$ and $l$ of the particles. By definition, the permutation $p=1$ is chosen
to correspond  to the order $1$, $2$, $3$ and $4$. In terms of
the Jacobi vectors, the kinetic energy $T$ is written as
\begin{equation}
  T=-{1\over M}\Bigl( \nabla^2_{\jacb_{1p}}
  +\nabla^2_{\jacb_{2p}}+\nabla^2_{\jacb_{3p}}\biggr)\ ,
\end{equation} 
where $M$ is the nucleon mass (hereafter $\hbar=c=1$).
For a given choice of the Jacobi vectors, the hyperspherical coordinates are
given by the so-called hyperradius $\rho$, defined by
\begin{equation}
   \rho=\sqrt{\jac_{1p}^2+\jac_{2p}^2+\jac_{3p}^2}\ ,\quad ({\rm independent\
    of\ }p)\ ,
    \label{eq:rho}
\end{equation}
and by a set of angular variables which in the Zernike and
Brinkman~\cite{zerni,F83} representation are (i) the polar angles $\hat
\jacb_{ip}\equiv (\theta_{ip},\phi_{ip})$  of each Jacobi vector, and (ii) two additional angles, called hyperangles,  $\hypfi_{2p}$ and $\hypfi_{3p}$ defined as
\begin{equation}
    \cos\hypfi_{2p} = \frac{ \jac_{2p} }{\sqrt{\jac_{1p}^2+\jac_{2p}^2}}\ ,
    \quad
    \cos\hypfi_{3p} = \frac{ \jac_{3p} }{\sqrt{\jac_{1p}^2+\jac_{2p}^2+\jac_{3p}^2}}\ ,
     \label{eq:phi}
\end{equation}
where $\jac_{jp}$ is the modulus of the Jacobi vector $\jacb_{jp}$. The set of angular
variables $\hat \jacb_{1p}, \hat \jacb_{2p}, \hat \jacb_{3p}, \hypfi_{2p}$, $\hypfi_{3p}$ is
denoted  hereafter as $\Omega_p$. The expression of a generic HH
function is
\begin{eqnarray}
 \lefteqn{ {\cal H}^{K,\Lambda, M}_{\ell_1,\ell_2,\ell_3, L_2 ,n_2,
     n_3}(\Omega_p) =\qquad\qquad} &&  \nonumber \\
  && {\cal N}^{\ell_1,\ell_2,\ell_3}_{ n_2, n_3} 
      \left [ \Bigl ( Y_{\ell_1}(\hat \jacb_{1p})
    Y_{\ell_2}(\hat \jacb_{2p}) \Bigr )_{L_2}  Y_{\ell_3}(\hat \jacb_{3p}) \right
    ]_{\Lambda M}  \nonumber \\
  && 
   \times (\sin\hypfi_{2p})^{\ell_1 }    (\cos\hypfi_{2p})^{\ell_2}
      P^{\ell_1+\frac{1}{2}, \ell_2+\frac{1}{2}}_{n_2}(\cos2\hypfi_{2p})
      \nonumber\\
      &&\times
         (\sin\hypfi_{3p})^{\ell_1+\ell_2+2n_2}
      (\cos\hypfi_{3p})^{\ell_3} 
      P^{\ell_1+\ell_2+2n_2+2, \ell_3+\frac{1}{2}}_{n_3}(\cos2\hypfi_{3p})\ ,
      \label{eq:hh4P}
\end{eqnarray}
where $P^{a,b}_n$ are Jacobi polynomials and the coefficients ${\cal
N}^{\ell_1,\ell_2,\ell_3}_{ n_2, n_3}$ normalization factors. The quantity 
$K=\ell_1+\ell_2+\ell_3+2(n_2+n_3) $ is the grand angular quantum
number.  The HH functions are the eigenfunctions of the hyperangular part of
the kinetic energy operator. Furthermore, 
$\rho^K   {\cal  H}^{K,\Lambda,M}_{\ell_1,\ell_2,\ell_3, L_2 ,n_2,
n_3}(\Omega_p)$ are homogeneous polynomials of the particle coordinates of
degree $K$.

A set of antisymmetric hyperangular--spin--isospin states of 
grand angular quantum number $K$, total orbital angular momentum $\Lambda$,
total spin $\Sigma$, and total isospin $T$  (for given values of
total angular momentum $J$ and parity $\pi$) can be constructed as follows:
\begin{equation}
  \Psi_{\mu}^{K\Lambda\Sigma T} = \sum_{p=1}^{12}
  \Phi_\mu^{K\Lambda\Sigma T}(i,j,k,l)\ ,
  \label{eq:PSI}
\end{equation}
where the sum is over the $12$ even permutations $p\equiv i,j,k,l$, and
\begin{eqnarray}
 \lefteqn{  \Phi^{K\Lambda\Sigma T}_{\mu}(i,j;k;l)
   =\qquad\qquad} &&  \nonumber \\
  && \biggl \{
   {\cal H}^{K,\Lambda}_{\ell_1,\ell_2,\ell_3, L_2 ,n_2, n_3}(\Omega_p)
      \biggl [\Bigl[\bigl( s_i s_j \bigr)_{S_a}
      s_k\Bigr]_{S_b} s_l  \biggr]_{\Sigma} \biggr \}_{JJ_z}
      \biggl [\Bigl[\bigl( t_i t_j \bigr)_{T_a}
      t_k\Bigr]_{T_b} t_l  \biggr]_{TT_z}\ .
     \label{eq:PHI}
\end{eqnarray}
Here, ${\cal H}^{K,\Lambda}_{\ell_1,\ell_2,\ell_3, L_2 ,n_2, n_3}(\Omega_p)$ is the
HH state defined in Eq.~(\ref{eq:hh4P}), and $s_i$ ($t_i$) denotes the spin 
(isospin) function of particle $i$. The total orbital angular  momentum $\Lambda$ of
the HH function is coupled to the total spin $\Sigma$ to give the total angular
momentum $JJ_z$, whereas $\pi=(-1)^{\ell_1+\ell_2+\ell_3} $. The
quantum number $T$ specifies the total isospin of the state. The
integer index $\mu$ labels the possible choices of hyperangular, spin and
isospin quantum numbers, namely
\begin{equation}
   \mu \equiv \{ \ell_1,\ell_2,\ell_3, L_2 ,n_2, n_3, S_a,S_b, T_a,T_b
   \}\ ,\label{eq:mu}
\end{equation}
compatibles with the given values of $K$, $\Lambda$, $\Sigma$, 
$T$, $J$ and $\pi$. Each state  $\Psi^{K\Lambda\Sigma T}_\mu$ entering 
the expansion of the 4N wave function must 
be antisymmetric under the exchange of any pair of particles. To this aim 
it is sufficient to consider states such that
\begin{equation}
    \Phi^{K\Lambda\Sigma T}_\mu(i,j;k;l)= 
    -\Phi^{K\Lambda\Sigma T}_\mu(j,i;k;l)\ ,
     \label{eq:exij}
\end{equation}
which is fulfilled when the condition
\begin{equation} 
    \ell_3+S_a+T_a = {\rm odd}\ , \label{eq:lsa}
\end{equation}
is satisfied. Note that many of the antisymmetric states $\Psi^{K\Lambda\Sigma T}_\mu$
are linearly dependent between themselves.

The 4N wave function can be finally written as
\begin{equation}\label{eq:PSI3}
  \Psi_C= \sum_{K\Lambda\Sigma T}\sum_{\mu} 
    u_{K\Lambda\Sigma T\mu}(\rho)
    \Psi_{\mu}^{K\Lambda\Sigma T}\ .,
\end{equation}
where the sum is restricted only to the linearly independent states.  
This expansion can be used to compute either a bound-state wave function or
the ``internal'' part of the scattering wave function (see next subsection). We have found convenient to expand the hyperradial functions
$u_{K\Lambda\Sigma T\mu}(\rho)$ in a 
complete set of functions, namely
\begin{equation}
     u_{K\Lambda\Sigma T\mu}(\rho)=\sum_{m=0}^{M-1} 
      c_{K\Lambda\Sigma T\mu m} \; g_m(\rho)
      \ ,     \label{eq:fllag}
\end{equation}
and we have chosen 
\begin{equation}
   g_m(\rho)= 
     \sqrt{b^{9}\frac{m!}{(m+8)!}}\,\,\,  
     L^{(8)}_m(b\rho)\,\,{\rm e}^{-\frac{b}{2}\rho} \ ,
      \label{eq:fllag2}
\end{equation}
where $L^{(8)}_m(b\rho)$ are Laguerre polynomials~\cite{abra} and 
$b$ is a parameter to be variationally optimized.

One of the problem we have to face is that the number of linearly independent states is still very high, and increases
noticeably with $K$. In order to
reduce the number of states to be included in the expansion, we adopt the same strategy as described in Refs.~\cite{Viv05,Viv20}.
Namely, we divide the basis in classes, depending on the value of the quantity $\mathcal{L} =\ell_1 + \ell_2 + \ell_3$
and the values of $n_2$, $n_3$, see Section~3 of Ref.~\cite{Viv05} for the detailed definition of the classes.
Hereafter, we consider only the definition of the classes of HH functions for the state having total angular momentum and parity $J^\pi=0^+$ and total isospin $T=0$, which is the case we are interested in. In fact, in the present work, we can safely disregard HH states with $T>0$.

Briefly,  the first two classes include the states with ${\cal L}=0$ and a selected set of states with ${\cal L}=2$,
the third class the remaining states with ${\cal L}=2$ and the fourth and the fifth classes the states with ${\cal L}=4$ and ${\cal L}=6$, respectively. We have found that the convergence of the various quantities depends critically on the value of ${\cal L}$. Classes with low values of ${\cal L}$, typically ${\cal L}=0,2$, require the inclusion of HH states with large values of $K$ whereas this is not the case for higher values of ${\cal L}$. The contributions of the fourth and fifth classes to either the binding energy or to scattering observables becomes smaller and smaller as $K$ is increased, in particular the contribution of the fifth class is practically negligible. This is due to the fact that states with large values of ${\cal L}$ suffer for a high centrifugal barrier describing with a low probability particles close to each other. This reduces the importance of the corresponding HH states (we remember that the nuclear force is short range). In the following, we report the results obtained using different basis sets of HH functions, 
each of one corresponding to different values of $K_\alpha$, $\alpha=1,5$.  These values specify that in the class $\alpha$, only
states of grand angular quantum number $K\leq K_\alpha$  are included. The values adopted in the present work are given in Table~\ref{tab:1},
together with the total number of HH functions included in the expansion.

The ground state is calculated using the expansion given in Eq.(\ref{eq:PSI3}), and expanding the
hyperradial functions as in Eq.(\ref{eq:fllag}). In order to describe with great accuracy this state the values of $K_\alpha$
given in the upper part of Table~\ref{tab:1} can be used. For the scattering state, the function given
in Eq.~(\ref{eq:PSI3}) is used to describe the internal part
of the wave function, namely the region where all the four nucleons are close to each other (the full wave function will be detailed in the next subsection).
In this case the expansion has to describe the transition between the internal region and the asymptotic region in which the clusters are well separated with the consequence that one needs to increase $K_{1-5}$. 
For both the ground state and the scattering states, we provide three basis sets with increasing values of $K_\alpha$, see Table~\ref{tab:1}.
The various basis sets will be used in Sect.~\ref{sec:res} to check the convergence of the results.

\begin{table}
\centering
  \caption{Different basis sets of HH functions used in this
  calculation. Each basis set is specified by giving the maximum grand angular quantum numbers for the various classes of HH states included in $\Psi_C$.
    The value of $K_\alpha$ for the class $\alpha$ means that we have included
    all HH functions with $K\leq K_\alpha$ (for the definition of the classes, see Refs.~\protect\cite{Viv05,Viv20}).
      }
    \label{tab:1}       % Give a unique label
  \begin{tabular}{l|ccccc|c}
    \hline\noalign{\smallskip}
    \multicolumn{7}{c}{For the ground state}\\
set & $K_1$ & $K_2$ & $K_3$ & $K_4$ & $K_5$ & $N$ \\
\noalign{\smallskip}\hline\noalign{\smallskip}
A1 & 28 & 20 & 20 & 20 & 0 & \np2,498 \\
A2 & 30 & 22 & 22 & 22 & 0 & \np3,145 \\
A3 & 32 & 24 & 24 & 24 & 0 & \np3,871 \\
\noalign{\smallskip}\hline
    \multicolumn{7}{c}{For the scattering states}\\
set & $K_1$ & $K_2$ & $K_3$ & $K_4$ & $K_5$ & $N$ \\
\noalign{\smallskip}\hline\noalign{\smallskip}
B1 & 46 & 42 & 32 & 40 & 20 & \np7,548 \\
B2 & 48 & 44 & 34 & 42 & 22 & \np8,838 \\
B3 & 50 & 46 & 36 & 44 & 24 & 10,053 \\
\noalign{\smallskip}\hline
  \end{tabular}
\end{table}

\subsection{The scattering wave function}
\label{sec:scatt}

In the following, a specific clusterization $A+B$ will be
denoted by the index $\gamma$. More specifically, $\gamma=1$ ($2$) stands for the
clusterization $p+\tri$ ($n+\het$). Let us consider a scattering state with total
angular momentum quantum number $JJ_z$, and parity $\pi$. Here, we are interested
only on the case where $J^\pi=0^+$, so in the following these values are understood. The wave function $\Psi^{\gamma L S}$ describing incoming clusters $\gamma$
with relative orbital angular momentum $L$ and channel spin $S$, coupled to $JJ_z$, can be written as
\begin{equation}
    \Psi^{\gamma LS}=\Psi_C^{\gamma LS}+\Psi_A^{\gamma LS} \ ,
    \label{eq:psica}
\end{equation}
where the core part $\Psi_C^{\gamma LS}$ describes the four particles when they
are close to each other; it can be conveniently expanded as in Eq.~(\ref{eq:PSI3}).
The other term, $\Psi_A^{\gamma LS}$, describes the relative motion of the two clusters in
the asymptotic regions, where the mutual interaction is
negligible (except for the long-range Coulomb interaction), and it can be decomposed
as a linear combination of the following functions
\begin{eqnarray}
  \Omega_{\gamma LS}^F &=& {1\over\sqrt{4}} {\cal A}\bigg\{
  \Bigl [ Y_{L}(\hat{\bm y}_\gamma)   [ \Phi_\gamma(ijk)  s_l]_{S} 
   \Bigr ]_{JJ_z}  {\frac{F_L(\eta_\gamma,q_\gamma y_\gamma)}{q_\gamma y_\gamma}}\bigg\}\ , \label{eq:psiof}  \\
  \Omega_{\gamma LS}^G &=& {1\over\sqrt{4}} {\cal A}\bigg\{
  \Bigl [ Y_{L}(\hat{\bm y}_\gamma)   [ \Phi_\gamma(ijk)  s_l]_{S}
   \Bigr ]_{JJ_z}  \frac{G_{L}(\eta_\gamma,q_\gamma y_\gamma)}{q_\gamma y_\gamma} (1-e^{-\beta y_\gamma})^{2L+1}\bigg\}   \ ,  \label{eq:psiog}
\end{eqnarray}
where $y_\gamma$ is the distance between the center-of-mass (c.m.) of clusters $A$
and $B$, $q_\gamma$ is the magnitude of the relative momentum between the
two clusters, and $\Phi_\gamma(ijk)$ are the bound state wave functions (clearly, $\Phi_1\equiv\Phi_\tri$ and
$\Phi_2\equiv\Phi_\het$).
In the present work, the trinucleon bound state wave functions $\Phi_\gamma(ijk)$ (for both $\het$ and $\tri$)
are described using the HH method~\cite{rep08,Nogga03}. Moreover,
the channel spin $S$ is
obtained coupling the angular momentum of the two clusters. In our
case, we have $S=0,1$. The symbol ${\cal A}$ means that the expression
between the curly braces has to be properly antisymmetrized, summing
over the permutations of the particles $(ijk),l$ with $l=1,\ldots,4$
($\Phi_\gamma(ijk)$ are already antisymmetric under the exchange of $ijk$).

The c.m. kinetic energy $E_\gamma$ in the channel $\gamma$ is defined by the relations
\begin{equation}
  E_T=-B(\tri)+E_1 = -B(\het)+E_2\, \label{eq:energy}
\end{equation}
where $E_T$ is the c.m. energy of the state and $B(\tri)$ and $B(\het)$ the
binding energies of $\tri$ and $\het$, respectively. Depending on the value
of $E_T$, $E_\gamma$ can be either positive or negative. In the present paper, we
are interested in the range of energies $-B(\tri)\leq E_T\leq -2B(d)$, where
$B(d)$ is the deuteron binding energy. Namely, we are below the
opening of the $d+d$ channel. When $E_\gamma>0$, the wave number $q_\gamma$ is defined as
\begin{equation}
  E_\gamma={q_\gamma^2\over 2\mu_\gamma}\ , \qquad
  {1\over \mu_\gamma} = {1\over M_A}+{1\over M_B} 
   \ ,\label{eq:tcm}
\end{equation}
and $M_X$ is the mass of the cluster $X$.
Clearly, in the case of a single nucleon $M_X=M$. In the present case $E_1$ is always positive, while
$E_2<0$ for $-B(\tri)\leq E_T\leq -B(\het)$ (see below to see how
this case has been treated). 

In Eqs.~(\ref{eq:psiof}) and~(\ref{eq:psiog}), the functions $F_L$ and
$G_{L}$ describe the asymptotic radial motion of the
clusters $A$ and $B$. If the two clusters are composed of $Z_A$ and $Z_B$
protons, respectively, the parameter $\eta_\gamma$ is defined as
$\eta_\gamma=\mu_\gamma Z_A Z_B e^2/q_\gamma$, where $e^2=1.43997$
MeV fm. The functions $F_L(\eta,qy)$ and $G_{L}(\eta,qy)$ are 
the regular and irregular Coulomb function, respectively. The term
$(1-e^{-\beta y_\gamma})^{2L+1}$ is used to ``regularize'' the 
irregular Coulomb function for $y_\gamma\rightarrow 0$
(see Ref.~\cite{Viv20} for more details), but it does not affect
the long range behavior. The parameter $\beta$ is usually chosen to be $\beta=0.25$ fm${}^{-1}$.
Let us also define
\begin{equation}
  \Omega_{\gamma LS}^\pm = \Omega_{\gamma LS}^G\pm  {\rm i}  \Omega_{\gamma LS}^F\ ,
  \label{eq:psiom}
\end{equation}
where $\Omega_{\gamma LS}^+$  ($\Omega_{\gamma LS}^-$) describes the
outgoing (ingoing) relative motion of the clusters specified by $\gamma$. In fact, their asymptotic behavious is described as
\begin{equation}
  G_{L}(\eta,qy)\pm {\rm i} F_L(\eta,qy) \rightarrow
  e^{\pm {\rm i} \bigl (q y-L\pi/2-\eta\ln(2qy)+\sigma_L\bigr ) }\ ,
\end{equation}
where $\sigma_L$ is the Coulomb phase shift.

If one of the clusters is a neutron (case $\gamma=2$), then $\eta=0$ and
the functions $F_L$ and $G_L$ reduce to
\begin{equation}
 {F_L(\eta,qy)\over qy } \rightarrow j_L(qy)\ , \qquad
 {G_L(\eta,qy)\over qy } \rightarrow - y_L(qy)\ , 
 \label{eq:eta0}
\end{equation}
where $j_L$ and $y_L$ are the regular and irregular spherical Bessel functions
defined, for example, in Ref.~\cite{abra}.
Finally, the general expression of $\Psi_A^{\gamma LS}$ entering
Eq.~(\ref{eq:psica}) is
\begin{equation}
  \Psi_A^{\gamma, L,S}= 
   \Omega_{\gamma, L,S}^F
  + \sum_{\gamma',L',S'} {\cal T}^{\gamma,\gamma'}_{L,S;L^\prime, S^\prime}
     \Omega_{\gamma', L^\prime, S^\prime }^+ \ ,
  \label{eq:psia}
\end{equation}
where the parameters ${\cal T}^{\gamma,\gamma'}_{LS,L^\prime  S^\prime}$ are the so-called $T$-matrix elements.  Of course, the sum over
$L^\prime$ and $S^\prime$ is over  all values compatible with the given $J$ and
parity $\pi$. In the present case, with $J^\pi=0^+$, we have simply $L,S=L',S'=0,0$.

Limiting ourselves to energies below the opening of the $d+d$ channels, the
$p+\tri$ scattering wave function in the $0^+$ state is given by
\begin{equation}
  \Psi^{1,0,0}(E_1) = \Psi_C^{1,0,0}+ \Omega_{1,0,0}^F  + {\cal T}^{1,1}_{0,0;0,0} \Omega_{1,0,0}^+
  + {\cal T}^{1,2}_{0,0;0,0} \Omega_{2,0,0}^+  \ .
  \label{eq:psia1}
\end{equation}
Here we have included the term $\Omega_{2,0,0}^+$ below the $n+\het$ threshold
even if the channel is energetically ``closed'' (namely the energy $E_2$, fixed by the relation $E_T =-B_{\het}+E_2$,
is such that $E_2<0$). In this case $q_2={\rm i}\alpha_2$, where $\alpha_2=\sqrt{2\mu_2 |E_2|}$, and $\Omega_{2,0,0}^+$ reduces to
\begin{equation}
  \Omega_{2,0,0}^+ = {1\over\sqrt{4}} {\cal A}\bigg\{
  \Bigl [ Y_{0}(\hat{\bm y}_\gamma)   [ \Phi_{2}(ijk)  s_l]_{0} 
   \Bigr ]_{0,0}  {e^{-\alpha_2 y_2}\over y_2}(1-e^{-\beta y_2})\bigg\}\ . \label{eq:psioe} 
\end{equation}
From a computational point of view, the presence of this term below the $n+\het$ threshold is very useful. In fact,
as $E_T\rightarrow -B(\het)$ (from below), $\alpha_2$ becomes rather small
and the component $\Omega_{2,0,0}^{+}$ will have a long-range tail, in spite of the
exponential term $e^{-\alpha_2 y_2}$.
Configurations of this type are rather difficult to be
constructed in terms of the ``internal'' part $\Psi_C$.  Therefore, 
including this term in the variational wave function
is decisive in order to solve the convergence problem found for $p+\tri$ scattering in Ref.~\cite{Viv20}
below the $n+\het$ threshold.

Above the $n+\het$ threshold, we have to consider also the wave function with the term $\Omega_{2,0,0}^F$, namely
\begin{equation}
  \Psi^{2,0,0}(E_2) = \Psi_C^{2,0,0}+ \Omega_{2,0,0}^F  + {\cal T}^{2,1}_{0,0;0,0} \Omega_{1,0,0}^+
  + {\cal T}^{2,2}_{0,0;0,0} \Omega_{2,0,0}^+  \ .
  \label{eq:psia2}
\end{equation}
We remember that the relation between $E_1$ and $E_2$ is given in Eq.~(\ref{eq:energy}).

\subsection{The monopole form factor}
\label{sec:monoff}
As discussed in the Introduction, the monopole form factor is extracted from the $e+\heq$ cross section. 
Below the $n+\het$ threshold, the process to be considered is $\heq(e,e'p)\tri$, while above
that threshold, the contribution of the process $\heq(e,e'n)\het$ has also  to be considered. However, for the
sake of simplicity, in the present subsection we work out the cross section
for the $\heq(e,e'p)\tri$ process only, giving the complete expression of $F_M(q)$ at the end. 
In any case, we restrict our study to energies below the $d+d$ threshold.

In the following $\bmk$ and $s$ ($\bmk'$ and $s'$) are the momentum
and spin projection of the incoming (outgoing) electron. The $\heq$ is considered at rest, while the final proton
($\tri$) has momentum $\bmp_1$ ($\bmp_3$) and spin projection $m_1$ ($m_3$). Clearly
$\bmq=\bmk-\bmk'$ and $\omega=k-k'$ are the momentum and energy transfer.
The electrons are ultrarelativistic, hence we will consider them as massless. For example,
in the MAMI experiment, the incident beam energy was in the range 430-780 MeV. As discussed above,
$\omega\approx 20$ MeV, so also the final electrons can be safely considered as ultrarelativistic. 

The cross section can be calculated using the Fermi Golden Rule
\begin{equation}
  d\sigma = {1\over 2} \sum_{ss'} \sum_{m_1,m_3} \sum_{\bmk',\bmp_1,\bmp_3} 2\pi \delta(E_i-E_f) |T_{fi}|^2 \delta_{\bmk,\bmk'+\bmp_1+\bmp_3}
  \ , \label{eq:cross1}
\end{equation}
where $E_i=k+M_4$ and
\begin{equation}
  E_f= k'+ M_3 + {\bmp_3^2\over 2M_3} + M +{\bmp_1^2\over 2M}\ ,\label{eq:ef}
\end{equation}
$M$, $M_3$, and $M_4$ being the nucleon, $\tri$, and $\heq$ masses, respectively.
In the previous expression, we can safely use the non relativistic expression
for the proton and $\tri$ kinetic energies.
The transition matrix element $T_{fi}$ is given by
\begin{equation}
  T_{fi} = 4\pi\alpha {H^\mu \ell_\mu\over (-Q^2)}\ ,\label{eq:tme}
\end{equation}
where $H^\mu$ and $\ell^\mu$ are the hadronic and leptonic matrix elements, respectively, and $Q^2=q^2-\omega^2$.
Above, $\alpha$ is the fine structure constant, $\alpha\approx 1/137$. The hadronic matrix
element is decomposed as usual in multipoles. We are interested to the transition $0^+_0\rightarrow 0^+_1$ induced by the electron scattering,
therefore, we only need to compute the  matrix element of the charge operator $\hat \rho(\bmq)$, which 
at leading order (or in impulse approximation)
is simply given by
\begin{equation}
  \hat\rho(\bmq) = \sum_{j=1}^4 f_p(q) {1+\tau_z(j)\over 2} e^{i\bmq\cdot\bmr_j}\ , \label{eq:rhoia}
\end{equation}
where $f_p(q)$ is the proton form factor chosen to be~\cite{Shen2012}
\begin{equation}
  f_p(q) = {1\over (1+ 0.056 \; q^2)^2} \ ,\label{eq:dff}
\end{equation}
$q$ being given in fm${}^{-1}$. 
In principle, the proton form factor should depend on $Q$, with $Q^2=q^2-\omega^2$, $\omega$ being the energy transfer. However,
in the process under consideration $\omega\approx 20$ MeV, while the typical values of $q$ are of the order of $1$ fm${}^{-1} \approx 200$ MeV,
so we can safely neglect $\omega^2$ with respect to $q^2$.

In the calculation, we will consider also the contribution of various corrections to the operator given in Eq.~(\ref{eq:rhoia}),
as relativistic corrections and meson-exchange terms,  derived within $\chi$EFT~\cite{Pastore13}. As we will see,
these latter contributions are small but sizable, in particular for large values of $q$. The wave function of the final $p+\tri$ state can be written as
\begin{eqnarray}
  \Psi^{1+3}_{m_3,m_1}(\bmp) &=& \sum_{SS_z LM JJ_z} ({1\over 2},m_3,{1\over 2},m_1|S,S_z)
  (L, M, S,S_z| J, J_z)\nonumber\\
  && \times 4\pi i^L e^{i\sigma_L} Y^*_{LM}(\hat\bmp) \Psi^{1,L,S}_{JJ_z}(E_1)\ , \label{eq:wftotal}
\end{eqnarray}
where $\bmp$ is the c.m. relative momentum between proton and $\tri$ in the final state,
$E_1=p^2/2\mu_1$, $\mu_1$ being the $p+\tri$ reduced mass, $\sigma_L$ is the Coulomb phase shift,
and $\Psi^{1,L,S}_{JJ_z}(E_1)$ the scattering wave function given in Eq.~(\ref{eq:psia1}).
Note that in this subsection, the dependence on $JJ_z$ is explicitly reported.
Now, as discussed above, we can reduce the calculation considering only the product $H^0\ell_0$ and the contribution
of the wave with $J=L=S=0$. Therefore
\begin{eqnarray}
  H^0 &=& \langle \Psi^{1+3}_{m_3,m_1}(\bmp)| \hat \rho(\bmq) | \Psi_0\rangle\ ,  \nonumber\\
  &\approx & \sqrt{4\pi}({1\over2},m_3,{1\over2},m_1|0,0) e^{i\sigma_0}
  \langle \Psi^{1,0,0}_{0,0}(E_1)| \hat \rho(\bmq) | \Psi_0\rangle \nonumber\\
  & = &\sqrt{4\pi}({1\over2},m_3,{1\over2},m_1|0,0) e^{i\sigma_0}
     \sqrt{4\pi} C_0^{000}(q,E_1)\ ,\label{eq:J0}
\end{eqnarray}
where $\Psi_0$ is the $\heq$ ground state wave function and $C^{000}_0(q,E_1)$ the monopole ($\ell=0$)
reduced matrix element (RME) of the charge operator, which we define as
\begin{equation}
  \langle \Psi^{1,0,0}_{0,0}| \hat \rho(\bmq) |\Psi_0\rangle=\sqrt{4\pi} C^{000}_{0}(q,E_1)\ .\label{eq:rme}
\end{equation}
Thus
\begin{equation}
  \sum_{m_3,m_1} H^0 (H^0)^* = (4\pi)^2 |C_0^{000}(q,E_1)|^2 \ ,
\end{equation}
The leptonic current matrix element $\ell^\mu$ is given by
\begin{equation}
\ell^\mu_{\bmk s,\bmk' s'} = \overline{u}(\bmk',s') \,\gamma^\mu \,u(\bmk,s) \ ,
\end{equation}
where we have chosen to normalize the four spinors as $u^\dagger u\,$=$\,1$. Clearly, we have to consider only $\ell^0$.
The sum over the electron spins can be now obtained easily
\begin{equation}
  \sum_{s,s'} \ell^0 (\ell^0)^* = 1 + \cos\theta \ ,
\end{equation}
where $\cos\theta=\hat k\cdot\hat k'$ and $\theta$ is the electron scattering angle. Thus
\begin{eqnarray}
  d\sigma &=& {1\over 2} \sum_{\bmk',\bmp_1,\bmp_3}\left({4\pi\alpha\over Q^2}\right)^2 (4\pi)^2 |C_0^{000}(q,E_1)|^2 (1+\cos\theta)\nonumber\\
  &&\times \delta\Bigl(k+M_4 - k' -  M - M_3 -{p_1^2\over 2M} - {p_3^2\over 2M_3}\Bigr)\delta_{\bmq,\bmp_1+\bmp_3}\ .
\end{eqnarray}
The total momentum of the two nuclear clusters is $\bmP=\bmp_1+\bmp_3$, while the relative momentum is
$\bmp=(M\bmp_3-M_3 \bmp_1)/(M+M_3)$. Clearly, $\sum_{\bmp_1,\bmp_3}\equiv \sum_{\bmP,\bmp}$, and
\begin{equation}
  {p_1^2\over 2M} + {p_3^2\over 2M_3} = {p^2\over 2\mu_1} + {P^2\over 2(M+M_3)}\equiv E_1+ {P^2\over 2(M+M_3)} \ .
\end{equation}
The momentum-conserving delta function fixes $\bmP=\bmq$. At this point we
can go to the continuum limit and write
\begin{equation}
  \sum_{\bmk',\bmp} = \int {d^3k'\over(2\pi)^3} {d^3p\over(2\pi)^3}\ .
\end{equation}
However, we will not integrate over the momentum direction of the outgoing electron since we want to compute the differential
cross section $d\sigma/d\hat k'\equiv d\sigma/d\Omega$. The modulus of $\bmk'$ is fixed by the energy-conservation delta.
We have
\begin{equation}
  \int dk'\; \delta\Bigl(k-\Delta B-k' - {q^2\over 2(M+M_3)}\Bigr) F(k')
  = f_R F(k')|_{k'=k-\Delta B+\cdots}\ , 
\end{equation}
where
\begin{equation}
    \qquad f_R={1\over 1+{k'-k\cos\theta\over (M+M_3)}}\ , \qquad \Delta B= E_1+ M+M_3-M_4\approx 20\ {\rm MeV}\ .
\end{equation}
The ``recoil'' factor $f_R$ can be safely approximated to $1$. For example, in the MAMI experiment we have  $k=795$ MeV (in the worst case)
and the outgoing detector is placed at an angle of $\theta=18,3^\circ$. Since $k'\approx k$, $f_R$ turns out to be $\approx 0.99$. So in the
following we will take simply $f_R=1$. Moreover $Q^2=(\bmk'-\bmk')^2-(k-k')^2= 4 k k' \sin^2(\theta/2)$.
Putting all together, we obtain
\begin{eqnarray}
  {d\sigma\over d\Omega} &\approx& {1\over 2} \int  {d^3p\over(2\pi)^3} {dk' k^{\prime 2}\over (2\pi)^3} \;
  \left({4\pi\alpha\over Q^2}\right)^2 (4\pi)^2 |C_0^{000}(q,E_1)|^2 (1+\cos\theta)\nonumber\\
  &&\times 2\pi \delta\Bigl(k-\Delta B - k' - {q^2\over 2(M+M_3)}\Bigr)\ ,\\ 
  &=& {4\over\pi} {\alpha^2\over Q^4} k^{\prime 2} (1+\cos\theta) \int d^3p\; |C_0^{000}(q,E_1)|^2\ , \\
  &=& {\alpha^2\over 4k^2\sin^4(\theta/2)} \cos^2(\theta/2) \int_0^\infty dE_1\; 8 \mu_1 p |C_0^{000}(q,E_1)|^2\ , \\
  &=& \left({d\sigma\over d\Omega}\right)_{Mott}  \int_0^\infty dE_1\;  8 \mu_1 p  |C_0^{000}(q,E_1)|^2 \ .
\end{eqnarray}
We remember that $p=\sqrt{2\mu_1 E_1}$ and $k'=k-\Delta B+\cdots$.
Above, ${d\sigma\over d\Omega}_{Mott}$ is the Mott cross section, i.e. the differential cross section
for the scattering of an electron by a point-like charge.

As in the experiment, we define
\begin{equation}
  |F_M(q)|^2 =  {d\sigma\over d\Omega}/ \Bigl[Z^2 4\pi \left({d\sigma\over d\Omega}\right)_{Mott}\Bigr]\ .
\end{equation}
Here $Z=2$ takes into account the total charge of the $\heq$ nucleus. Finally
\begin{equation}
  |F_M(q)|^2 = {1\over 16\pi} \int_0^\infty dE_1\; 8p\mu_1 |C_0^{000}(q,E_1)|^2\ . \label{eq:fmq2}
\end{equation}
This expression is correct up to the $n+\het$ threshold. Above, it has to be modified as
\begin{equation}
  |F_M(q)|^2 = {1\over 16\pi} \sum_{\gamma=1,2}\int_0^\infty dE_\gamma\; 8p_\gamma\mu_\gamma |C_0^{000}(q,E_\gamma)|^2\ , \label{eq:fmq2b}
\end{equation}
where $p_\gamma=\sqrt{2\mu_\gamma E_\gamma}$, $\mu_\gamma$ is the reduced mass for the clusterization $\gamma$, and $C_0^{000}(q,E_\gamma)$
the RME coming from  $\langle \Psi^{\gamma,0,0}_{0,0}(E_\gamma)| \hat \rho(\bmq) | \Psi_0\rangle $.
The expression given in Eq.~(\ref{eq:fmq2}) agrees with those used in Ref.~\cite{Bacca13}
(where the sum over the final states was obtained using the LIT), as discussed in Appendix~\ref{app:a}.

\section{Results}
\label{sec:res}

First of all, we compute the scattering wave functions $\Psi^{\gamma,0,0}_{0,0}(E_\gamma)$ for various center-of-mass (c.m.) energy $E_\gamma$. 
The calculation has been performed by increasing the size of the HH basis as discussed
before. As an example, we report the results for the $p+\tri$ phase-shift calculated at $E_1=0.20$, $0.55$, $0.74$, and $1.50$ MeV in
Table~\ref{tab:ps-conv}. The calculations have been performed with the N3LO500/N2LO500 and N3LO600/N2LO600 NN+3N chiral interactions.
As it can be seen, the convergence is quite slow especially at the lowest energy, namely close to the resonance position.
As the energy is increased, the convergence is faster. At $E=1.50$ MeV (above the opening of the $n+\het$ channel),
the convergence is almost achieved. In the last line of the table, the extrapolated value for the phase-shift
as obtained from the calculated values for the different basis set is also reported (see Appendix~\ref{app:b} for more details about the
extrapolation procedure).

\begin{table}
\centering
  \caption{ Convergence of the $p+\tri$ phase-shift (deg) calculated at four different energies $E_1$,
    for the three different basis sets specified in Table~\protect\ref{tab:1}.
    The calculations have been performed with the N3LO500/N2LO500 and N3LO600/N2LO600 interactions.
    At $E_1=0.74$ MeV ($E_1=0.73$ MeV) for N3LO500/N2LO500 (N3LO600/N2LO600), we are just below the $n+\het$ threshold.
    In the lines labeled ``Extr.'', 
    the extrapolated values for the phase shifts are reported. 
      }
    \label{tab:ps-conv}       % Give a unique label
  \begin{tabular}{l|cccc}
    \hline\noalign{\smallskip}
    &\multicolumn{4}{c}{N3LO500/N2LO500}\\
basis set & $E_1=0.20$ MeV & $E_1=0.55$ MeV & $E_1=0.74$ MeV & $E_1=1.50$ MeV \\
\noalign{\smallskip}\hline\noalign{\smallskip}
B1 & $46.0$ & $82.2$ & $100.2$ & $86.6$ \\
B2 & $46.7$ & $82.6$ & $100.4$ & $86.8$ \\
B3 & $47.3$ & $82.8$ & $100.5$ & $86.9$ \\
\noalign{\smallskip}\hline
Extr. & $49.2$ & $83.6$ & $101.0$ & $87.2$ \\
\noalign{\smallskip}\hline
    &\multicolumn{4}{c}{N3LO600/N2LO600}\\
basis set & $E_1=0.20$ MeV & $E_1=0.55$ MeV & $E_1=0.73$ MeV & $E_1=1.50$ MeV \\
\noalign{\smallskip}\hline\noalign{\smallskip}
B1 & $31.4$ & $70.3$ & $91.0$ & $82.1$ \\
B2 & $32.0$ & $70.9$ & $91.4$ & $82.2$ \\
B3 & $32.5$ & $71.4$ & $91.7$ & $82.3$ \\
\noalign{\smallskip}\hline
Extr. & $34.5$ & $73.2$ & $92.9$ & $82.6$ \\
\noalign{\smallskip}\hline
  \end{tabular}
\end{table}

\begin{figure}
\centering
  \includegraphics[width=0.8\textwidth,clip,angle=0]{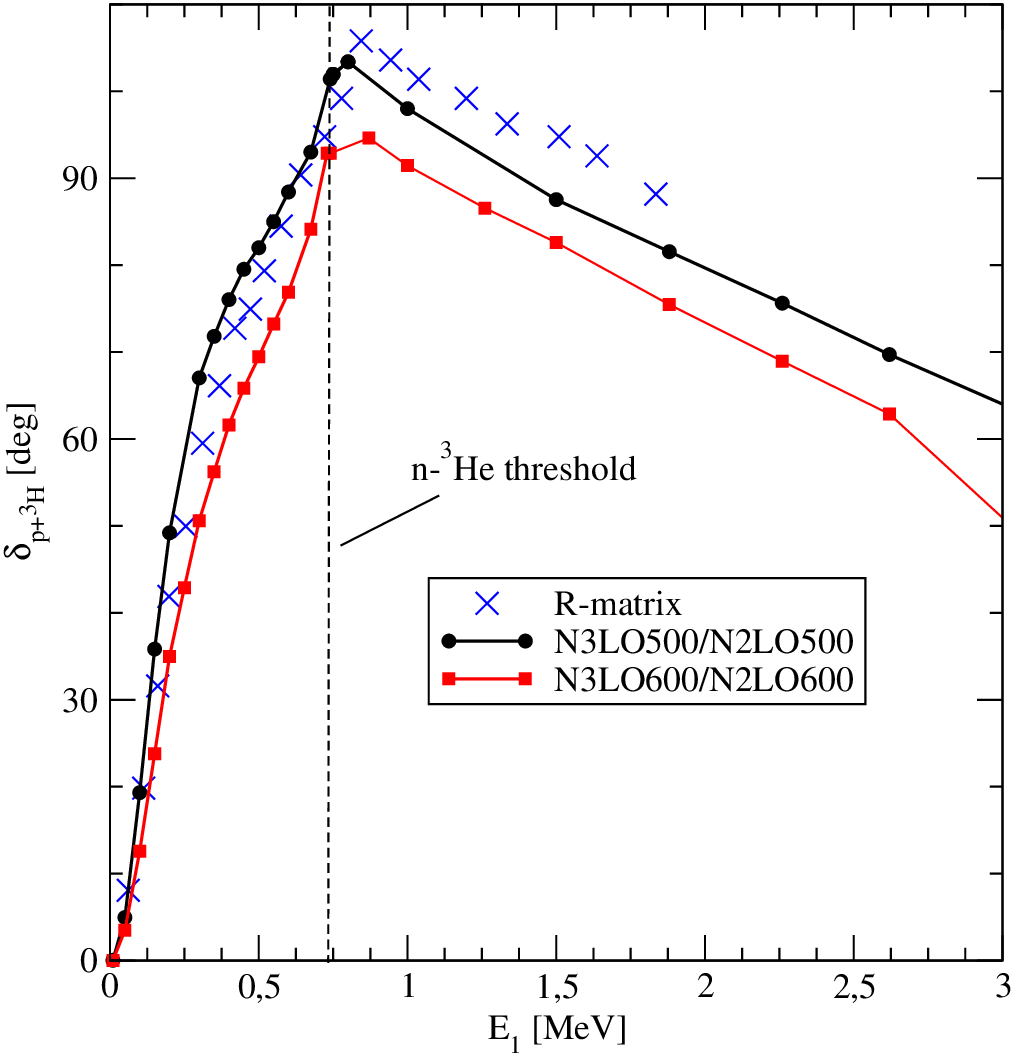}
\caption{(color online)  $0^+$ $p+\tri$ phase shift as function
      of the c.m. kinetic energy $E_1$ calculated
      with the NN+3N N3LO500/N2LO500 (black curve) and N3LO600/N2LO600 (red curve) interactions. Crosses: phase-shift
    extracted from the R-matrix analysis~\cite{HH08}. The vertical dashed line 
    denotes the energy of the $n+\het$ threshold. }
\label{fig:1}       % Give a unique label
\end{figure}

In Fig.~\ref{fig:1}, we report the ``extrapolated'' phase-shifts calculated
with both interactions, compared with the available ``experimental'' results,
extracted from an R-matrix analysis performed in Ref.~\cite{HH08}.
Below the $n+\het$ threshold, the phase shifts show the typical resonance behavior: a sharp increase
with the phase shift reaching the value $90$ deg. The convergence pattern is similar for both
interactions. The position of the resonance can be deduced by looking to the so-called
time delay, namely the quantity $\delta'(E_1)=d\delta(E_1)/dE_1$. The energy $E_R$ of the resonance is deduced for the energy $E_1$ where
$\delta'(E_1)$ has a maximum and its width $\Gamma$ by the value of $2/\delta'(E_R)$. In the present case, we obtain
$E_R\approx 0.1$ MeV and $\Gamma\approx 0.4$ MeV.

By inspecting Fig.~\ref{fig:1}, it can be noticed that above the $n+\het$ threshold there is a sharp change, since the phase shift
starts to decrease steadily. This behavior is also observed in the ``experimental'' phase shifts. Above the $n+\het$ threshold,
evidently the dynamics of the reaction changes. In fact, the process $p+\tri\rightarrow n+\het$  starts to become rapidly dominant,
and the reaction does not show any sign of the production of the resonant state.

\begin{figure}
\centering
  \includegraphics[width=0.8\textwidth,clip,angle=0]{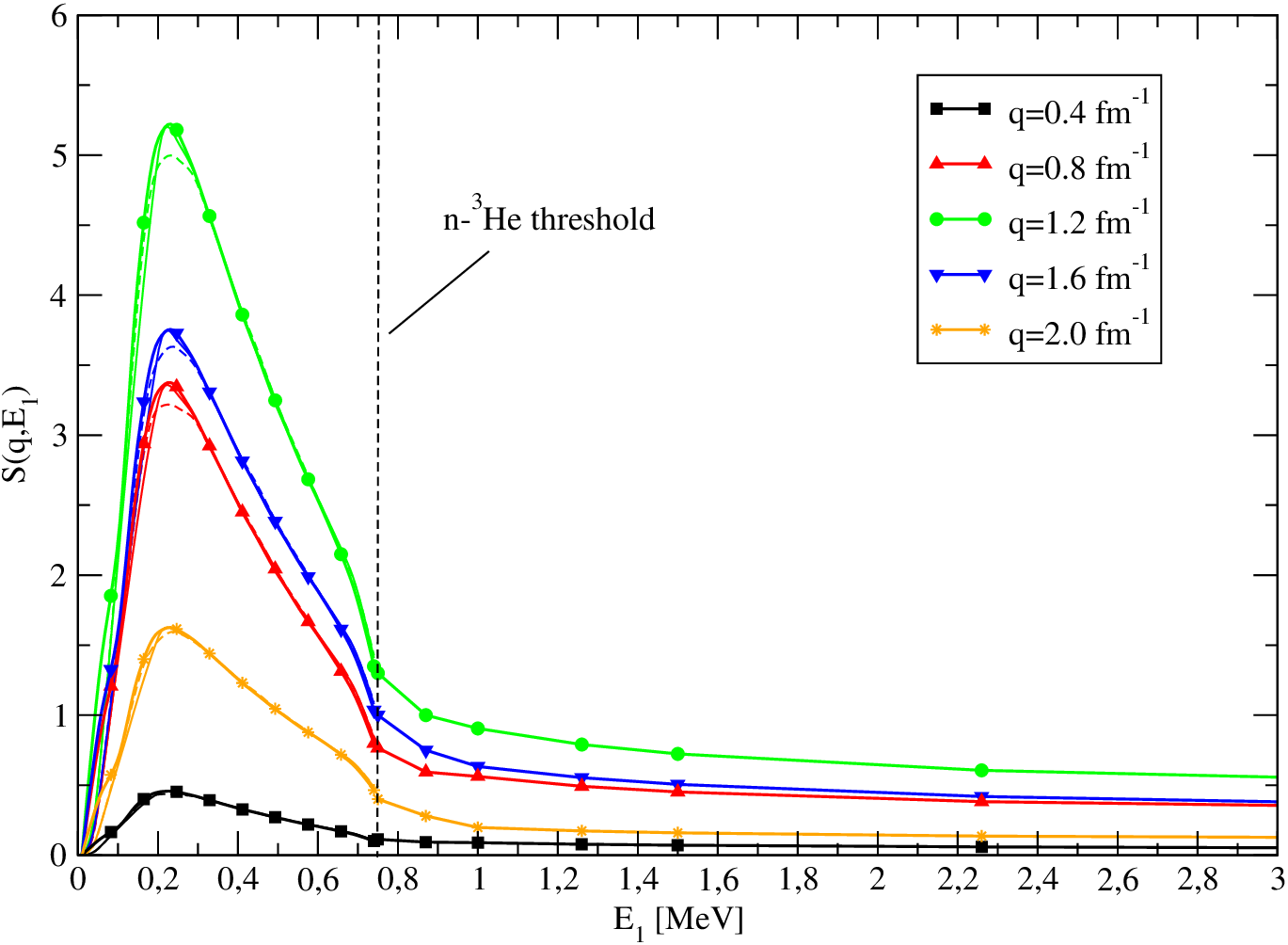}
\caption{(color online) The function $S(q,E_1)$ calculated for various values of $q$ and
  with the N3LO500/N2LO500 interaction as function of the c.m. kinetic energy $E_1$ of the $p+\tri$ clusters.
  The vertical line denotes the opening of the $n+\het$ channel. For each value of $q$, the dashed,  thin solid, and thick solid lines are obtained
  using the sets B1, B2, and B3 in the expansion of the scattering wave function, respectively. The calculations with set B2 are
  practically coincident with the results obtained with set B3 and they can be hardly distinguished.
  Above the $n+\het$ threshold only the calculations with set B3
  are reported, as the results obtained with sets B1 and B2 are practically coincident.
  Furthermore, the results are very independent on the basis set A1, A2, or A3 used to describe the ground state wave function.}
\label{fig:sqo1}       % Give a unique label
\end{figure}

Using the so determined wave functions, we can calculate the matrix elements with the $\heq$ ground state wave function
using Eq.~(\ref{eq:J0}). In Fig.~\ref{fig:sqo1}, we report the function $S(q,E_1)=\sum_{\gamma=1,2} 8p\mu_\gamma |C_0^{000}(q,E_\gamma)|^2$ vs $E_1$
calculated with the N3LO500/N2LO500 interaction (we remember that $E_1$ and $E_2$ are related by Eq.~(\ref{eq:energy})).
Clearly, for $E_1<E_{thr}$, where $E_{thr}$ is the energy where the $n+\het$ channel opens, $S(q,E_1)=8p\mu_1 |C_0^{000}(q,E_1)|^2$.

By inspecting Fig.~\ref{fig:sqo1}, first of all, we can note the good convergence reached by this quantity as the basis sets
used to describe the wave functions are enlarged. In fact, the results are very independent on the basis set A1, A2, or A3 used
to describe the ground state wave function. Regarding the scattering wave functions, $S(q,E_1)$ comes out to be practically
the same using basis sets B2 and B3.

\begin{figure}
\centering
  \includegraphics[width=0.8\textwidth,clip,angle=0]{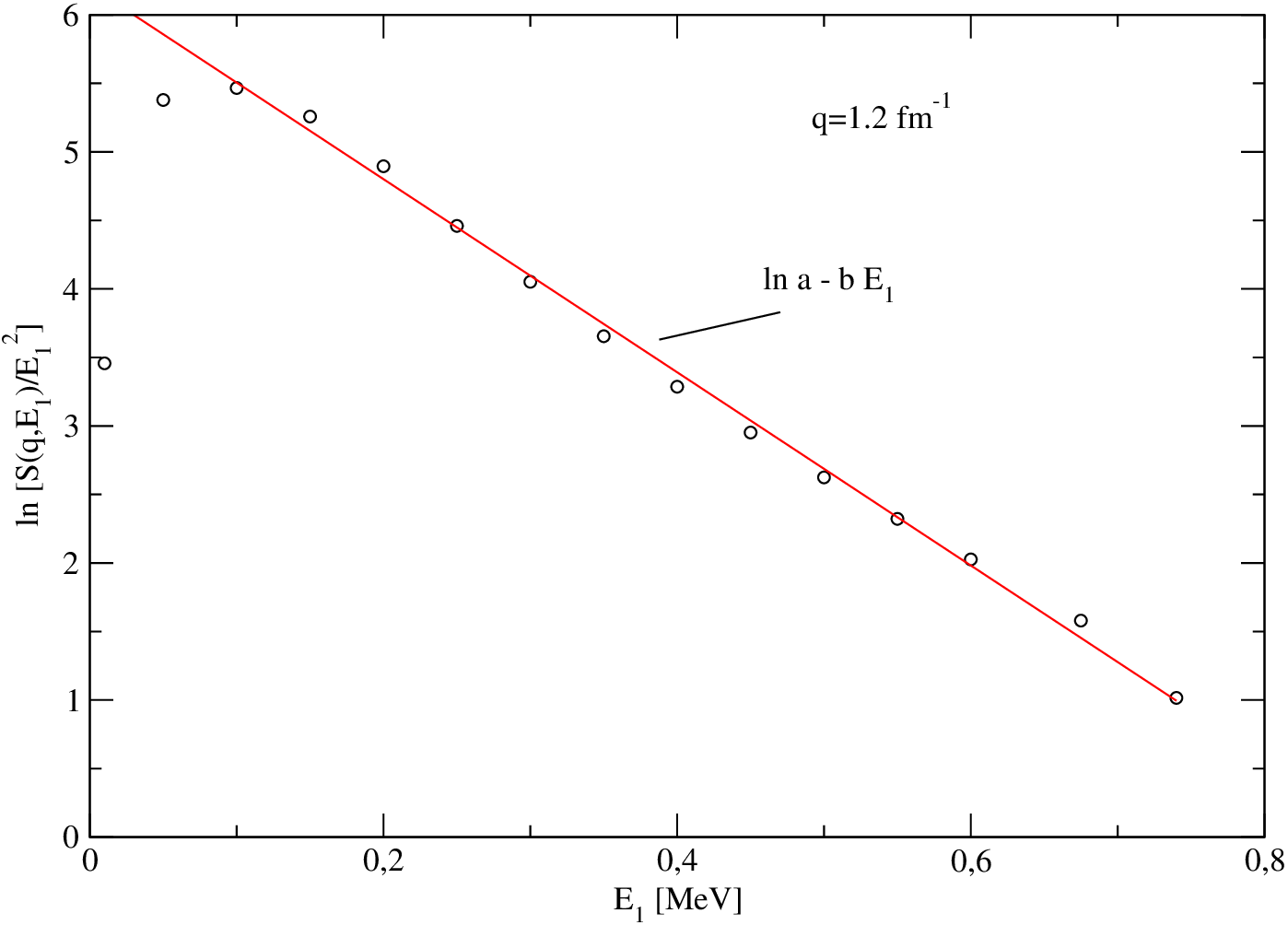}
  \caption{(color online) Example of the fit of the calculated $S(q,E_1)$ with the function $a E_1^2 e^{-b E_1}$.
    Solid circles: $\ln[S(q,E_1)/E_1^2]$ calculated for $q=1.2$ fm${}^{-1}$ with the N3LO500/N2LO500 interaction and the
    set B3 of HH functions. At it can be seen, for $E_1>0.2$ MeV the calculated values of this function are
    located in a very good approximation along a line, which is then fitted using a least-square method
    using the function $\ln(a)-bE_1$.}
  \label{fig:fit}       % Give a unique label
\end{figure}

\begin{figure}
\centering
  \includegraphics[width=0.8\textwidth,clip,angle=0]{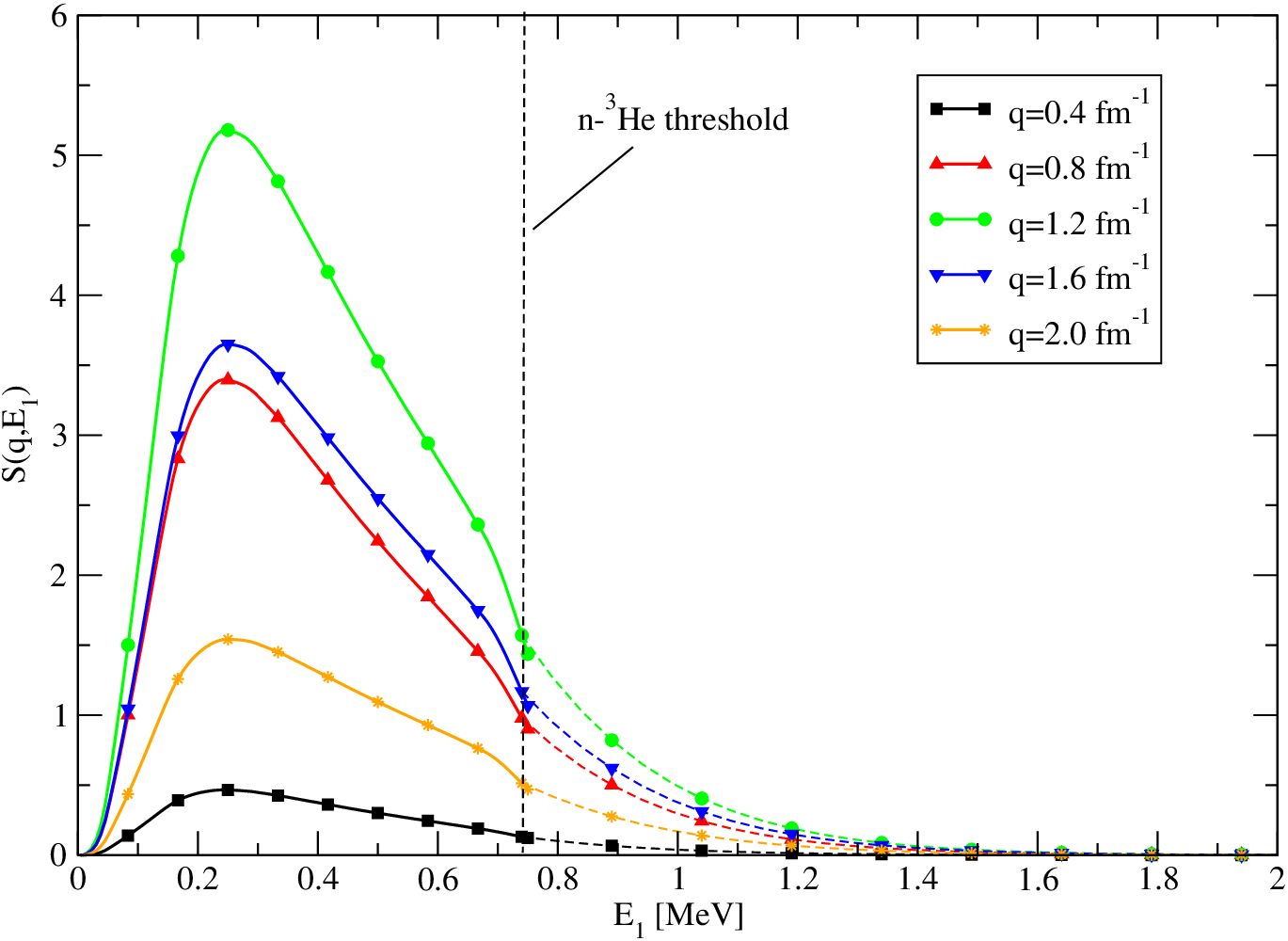}
  \caption{(color online) The same as in Fig.~\protect\ref{fig:sqo1}. The dashed lines for $E_1>E_{tr}$ show 
    the functions $a E_1^2 e^{-b E_1}$, fitted as explained in the main text. Only the calculation of set B3 of the HH basis is
    shown.  }
  \label{fig:sqo2}       % Give a unique label
\end{figure}

Second, we note that $S(q,E_1)$ has a form of a peak, related to the formation of the $0^+_{1}$ state in the scattering
process. However, above $E_{thr}$, the energy dependence of $S(q,E_1)$ becomes totally different. For such energies, 
the process is dominated by the direct charge-exchange reaction $p+\tri\rightarrow n+\het$, and 
the dynamics is not anymore related to the excitation of the $0^+_1$  state. We have then adopted the following procedure.
\begin{enumerate}
\item We have fitted the calculated $S(q,E_1)$ for $0<E_1<E_{thr}$ with the function $a E_1^2 e^{-b E_1}$, determining the
  parameters $a$ and $b$ using a least-square method.
  \item We have modified Eq.~(\ref{eq:fmq2}) as
\begin{eqnarray}
  |F_M(q)|^2 &=& {1\over 16\pi} \Bigl[\int_0^{E_{thr}} dE_1\; S(q,E_1) +\int_{E_{thr}}^\infty dE_1 \; a E_1^2 e^{-b E_1}\Bigr]\ ,\nonumber\\
  &=& {1\over 16\pi} \Bigl[\int_0^{E_{thr}} dE_1\; S(q,E_1) +{a\over b^3} \bigl(2+b E_{thr}(2+b E_{thr})\Bigr) e^{-b E_{thr}} \Bigr]\ .
  \label{eq:fmq2tail}
\end{eqnarray}
\end{enumerate}
In Fig.~\ref{fig:fit}, we show an example of the fit, plotting the function  $\ln\Bigl(S(q,E_1)/E_1^2\Bigr)$.
As it can be seen, for $E_1>0.20$ MeV the calculated values of this function are located in a very good approximation along a line,
easily fitted by the chosen function $\ln(a)-b E_1$.  We have then prolonged $S(q,E_1)$ above $E_{thr}$ with
the simple function $aE_1^2 e^{-b E_1}$. In Fig.~\ref{fig:sqo2}, we report the functions $S(q,E_1)$
as in Fig.~\ref{fig:sqo1}, but for $E_1>E_{thr}$ they are continued with the results of the fit (thin lines for $E_1>E_{thr}$).
The contribution of the $E_1>E_{thr}$ region to the integral turns out to be  about 10\%.
At the end, $|F_M(q)|^2$ is calculated using solely $S(q,E_1)$ determined for $E_1<E_{thr}$.

\begin{figure}
\centering
  \includegraphics[width=0.8\textwidth,clip,angle=0]{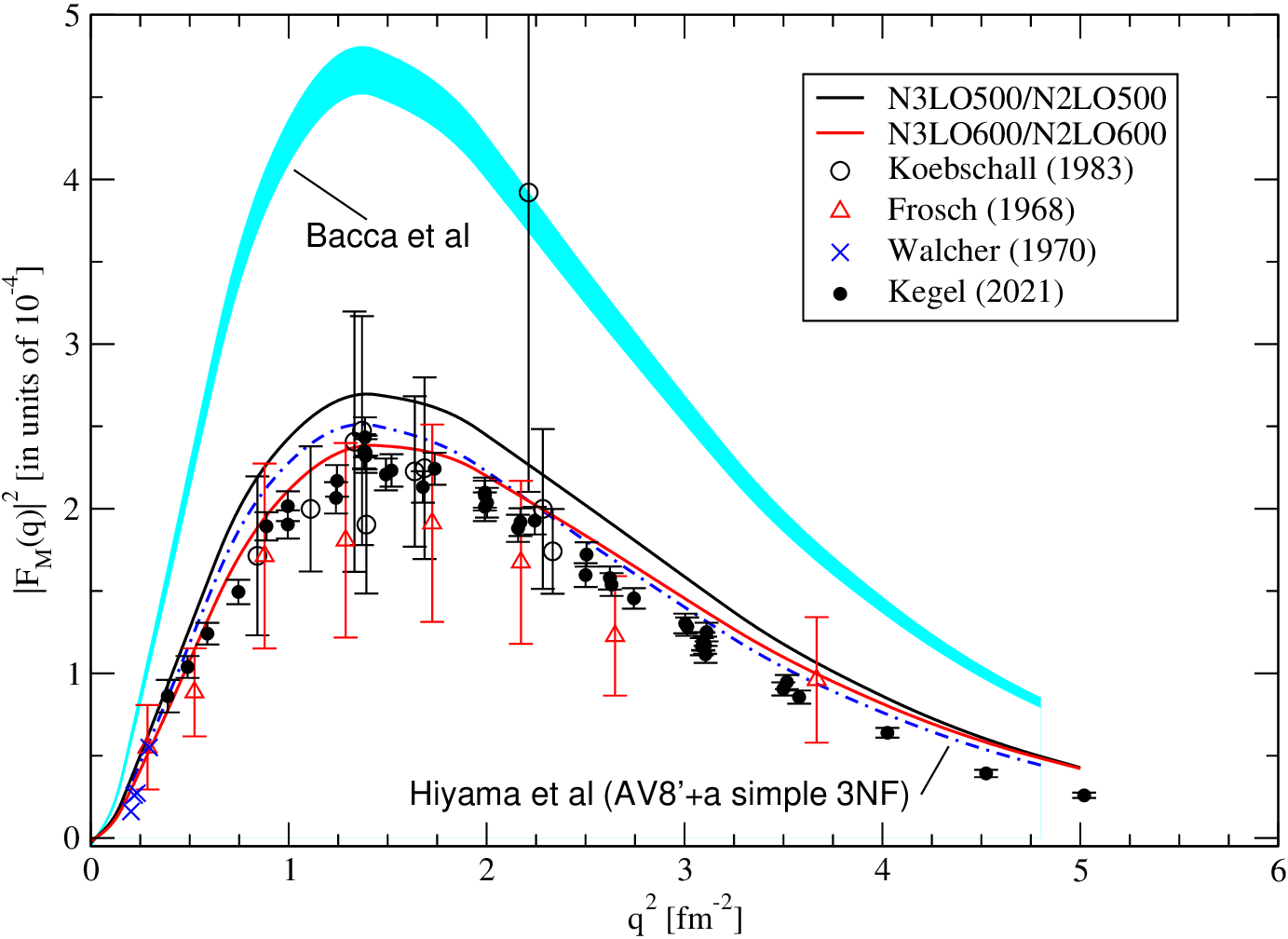}
  \caption{(color online) The monopole form factor $|F_M(q)|^2$ calculated using Eq.~(\protect\ref{eq:fmq2tail}) as function of $q$,
    obtained with the N3LO500/N2LO500 and N3LO600/N2LO600 interactions. The charge operator includes only the
    one-body leading-order (impulse approximation) term given in Eq.~(\protect\ref{eq:rhoia}). 
    The experimental data are from Refs.~\protect\cite{W70,Fea65,Kea83,Kegel23}
    The results of the calculations of Refs.~\protect\cite{Hiyama04,Bacca13} are also reported. }
  \label{fig:fmq2}       % Give a unique label
\end{figure}

The result of Eq.~(\ref{eq:fmq2tail}) is reported in Fig.~\ref{fig:fmq2}, for the two adopted interactions,
including only the leading order (or impulse approximation) contribution.
The calculations are compared with the experimental data of
Refs.~\cite{W70,Fea65,Kea83,Kegel23} and with some of the theoretical values reported in the literature~\cite{Hiyama04,Bacca13}.
As it can be seen, in this case the calculations are in reasonable
agreement with the experimental data, in particular with the MAMI data~\cite{Kegel23}. The spread between the calculations
obtained with the N3LO500/N2LO500 and N3LO600/N2LO600, related to the different cutoff values used to regularize the potentials,
reflects our current ignorance about the short-range part of the nuclear interaction.

\begin{figure}
\centering
  \includegraphics[width=0.8\textwidth,clip,angle=0]{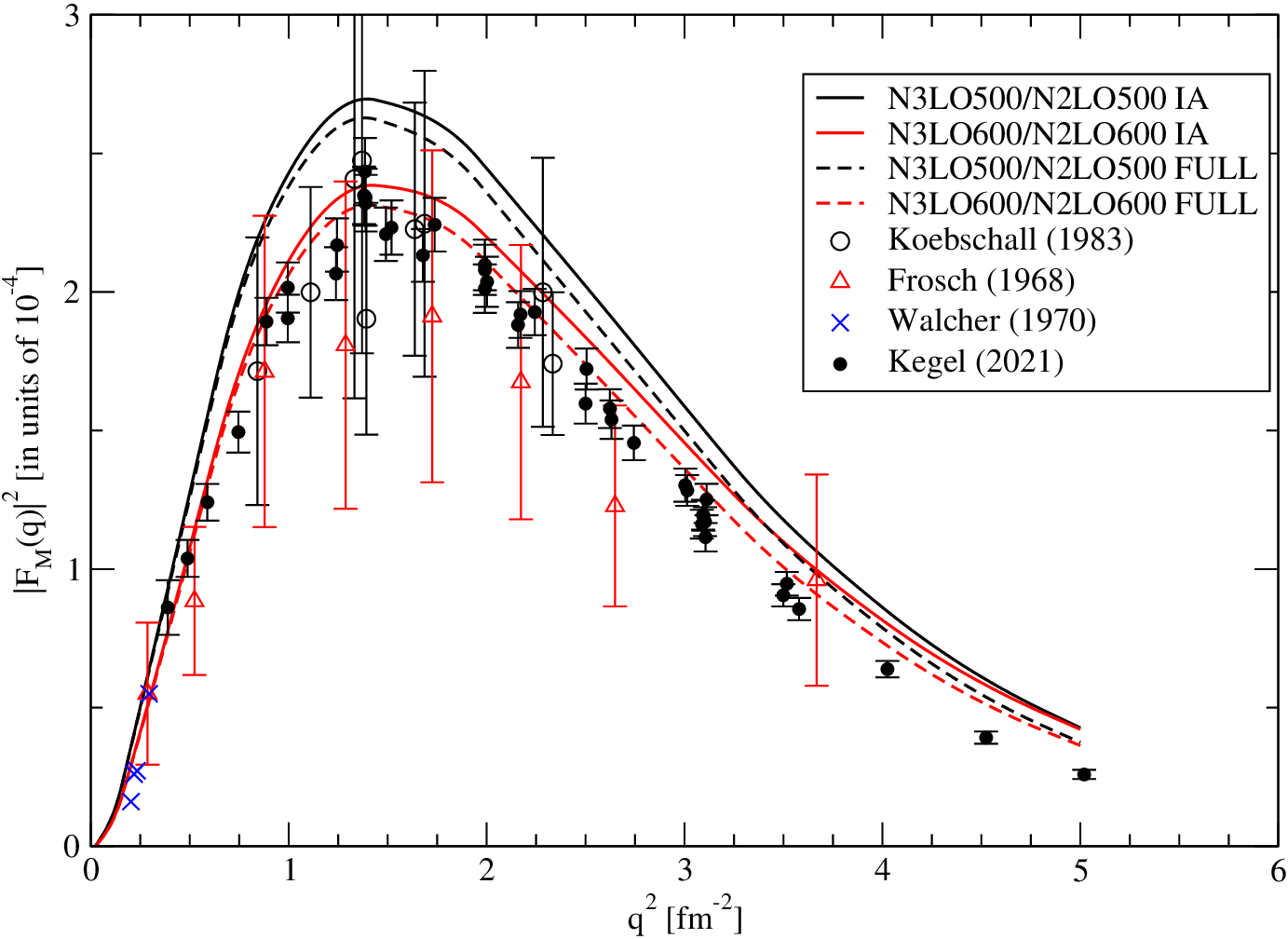}
  \caption{(color online) The same as in Fig.~\protect\ref{fig:fmq2} but adding the results obtained with the
    full charge operator (dashed lines), as derived in Ref.~\protect\cite{Pastore13}. }
  \label{fig:fmq2mec}       % Give a unique label
\end{figure}

Finally, in Fig.~\ref{fig:fmq2mec}, we report a calculation where the corrections beyond the leading order term for the
nuclear charge operator (the relativistic corrections and meson-exchange terms) are included (``full'' calculation).
As it can be seen, the ``full'' calculations slightly reduces the monopole form factor, especially for large values
of $q$, bringing the calculations very close to the MAMI data.

\section{Conclusion}
\label{sec:conc}

In this paper, we have studied the $\heq$ monopole form factor by calculating the
$\heq(e,e'p)\tri$ and $\heq(e,e',n)\het$ cross sections. The monopole form factor (squared) is then defined by the ratio of this
cross section with the Mott cross section. This procedure requires the summation over all possible final energies.
However, above the opening of the $n+\het$ threshold, the dynamics
of the process appears to drastically change, not involving anymore the formation of the $\heq$ first excited state (that
in this approach enters as a resonant state). We have then calculated the response $S(q,E_1)$ up to $E_{tr}$, using an approximate procedure to
complete the summation for $E>E_{tr}$. A similar procedure has been adopted also in the MAMI experiment~\cite{Kegel23}, as the contribution
of the process $0^+_0\rightarrow 0^+_1$ has to be extracted in some way from the measured cross section of the process $\heq(e,e'p)\tri$,
(see Ref.~\cite{Kegel23} for the procedure adopted in the MAMI experiment).

As it can be seen by inspecting Fig.~\ref{fig:fmq2}, our results are in reasonable agreement with the data, and 
with the theoretical study by Hiyama {\it et al.}~\cite{Hiyama04} and the more recent calculations reported in Refs.~\cite{Michel23,M23}.
However, they are at variance with the calculation of Ref.~\cite{Bacca13}. 

Although limited to just two nuclear interaction models derived within the $\chi$EFT framework, we believe that the results obtained are significant and representative. However, it is essential to study this observable also using other interaction models. We plan to perform this study in the future, 
in order to better understand how the
calculated $F_M(q)$ depends on the interaction. 

\begin{appendices}

\section{The formulation of Ref.~\cite{Bacca13}}
\label{app:a}
In Ref.~\cite{Bacca13}, the monopole form factor is defined as
\begin{equation}
  |\tilde F_M(q)|^2 = {1\over Z^2} \int d\omega {\cal S}(q,\omega)\ ,
\end{equation}
where $q,\omega$ are as usual the momentum and energy transfer. The quantity ${\cal S}(q,\omega)$ is defined as
\begin{equation}
  {\cal S}(q,\omega) = \sum_n |\langle n |{\cal M}(q)|0\rangle|^2 \delta(\omega-E_n+E_0) \ .
\end{equation}
where $|0\rangle$ and $|n\rangle$ are eigenfunctions of the 
the nuclear Hamiltonian, $E_0$ and $E_n$ the corresponding eigenvalues, and
\begin{equation}
  {\cal M}(q) = {f_p(q)\over 2}\sum_{j=1}^4 j_0(q r_j) \ ,
\end{equation}
is the isoscalar monopole operator ($j_0$ is a Bessel function). Note that this operator is the $\ell=0$
isoscalar component of our operator $\hat\rho(\bmq)$ given in Eq.~(\ref{eq:rhoia}). In our case $|0\rangle$ is the $\heq$
ground state and $|n\rangle$ the $p+\tri$ scattering state, given in Eq.~(\ref{eq:wftotal})
(here we limit ourselves to consider states below the $n+\het$ threshold). Considering only the
contribution of the $L=S=J=0$ wave, we have
\begin{eqnarray}
  {\cal S}(q,\omega) &=& \sum_{m_1,m_3,\bmp} ({1\over2},m_3,{1\over 2},m_1|0,0)^2 4\pi
  |\langle \Psi^{1,0,0}_{0,0}| {\cal M}(q) |\Psi_0\rangle|^2\nonumber\\
  &&\qquad \times \delta(\omega-M-M_3-{p^2\over2\mu_1}+M_4)\ ,
\end{eqnarray}
where, as in subsection~\ref{sec:monoff}, we can neglect the recoil term $q^2\over 2(M+M_3)$ in the $\delta$-function.
Introducing the RME as in Eq.~(\ref{eq:rme}) and changing $\sum_{\bmp}\rightarrow \int d^3p/(2\pi)^3$, we have
\begin{equation}
  {\cal S}(q,\omega) = \int {d^3p\over (2\pi)^3}\; ( 4\pi)^2 |C^{000}_{0}(q,E_1)|^2 
  \delta(\omega-M-M_3-{p^2\over2\mu_1}+M_4)\ .
\end{equation}
The integration over $d\hat p$ gives simply $4\pi$, and changing integration variable to $E_1={p^2\over2\mu_1}$, we obtain
\begin{equation}
  {\cal S}(q,\omega) = \int dE_1\; 8p\mu_1 |C^{000}_{0}(q,E_1)|^2  \delta(\omega-M-M_3-{p^2\over2\mu_1}+M_4)\ .
\end{equation}
Finally, by integrating over $\omega$, we have
\begin{equation}
  |\tilde F_M(q)|^2 = {1\over 4}  \int dE_1\; 8p\mu_1 |C^{000}_{0}(q,E_1)|^2\ .
\end{equation}
Note that in Ref.~\cite{Bacca13}, the monopole form factor which is compared with the data is
$|\tilde F_M(q)|^2/4\pi$, which therefore agrees with our definition of $|F_M(q)|^2$ given in Eq.~(\ref{eq:fmq2}).

\section{The extrapolation procedure}
    \label{app:b}

    In this appendix, we present the procedure adopted to extrapolate a given quantity $X$ (for example, the $p-\tri$ phase-shift $\delta$, etc.)
    estimating the ``missing'' part caused by the truncation of the HH basis in the calculations. 
    First of all, we perform several calculations of $X$
    using different basis sets $n=0,\ldots,N$ of HH functions (usually, $N=6$). Basis set $n=0$ is characterized by
    a given choice of the grand angular quantum numbers $K_\alpha^{(0)}$ for each class $\alpha=1,\ldots,5$.
    Then, for the basis set $n$, $K_\alpha=K_\alpha^{(0)}+2n$, etc.
    Let us denote with $X_n$ the quantity $X$ calculated with the basis set $n$. We have found that
    the ratios
    \begin{equation}
      x_n={X_n-X_{n-1}\over X_{n-1}-X_{n-2}}\ , \qquad n=2,\ldots ,N\ ,
    \end{equation}
    are, in a good approximation, independent of $n$. Namely, $x_n\approx x$, and $x$ is always less than $1$.
    That means, that each time the $K_\alpha$ are increased by $2$, the ``increment'' of $X$ is reduced by a factor $x$. 
    Therefore,    assuming that this property is maintained also for $n>N$, we can extrapolate the final value $X_\infty$
    of $X$ for an ``infinite'' basis, as
    \begin{equation}
      X_\infty = X_N + x \Delta + x^2 \Delta + x^3 \Delta +\cdots = X_N+{x\over 1-x} \Delta\ ,
      \label{eq:conv}
    \end{equation}
    where $\Delta = X_N-X_{N-1}$. Typical values of $x$ are around $0.8$.  In Table~\ref{tab:app},
    we report an example of this procedure. Finally, we estimate the ``error'' of $X_\infty$ by varying $x$ by $\pm2.5$\%.

    \begin{table}
\centering
      
    \caption{Convergence of the $p+\tri$ phase-shift $\delta$ at $E_1=0.20$ MeV,
      calculated with the N3LO500/N2LO500 interaction, using basis set $n$.
      In the third column, the ratio $x_n=(X_n-X_{n-1})/(X_{n-1}-X_{n-2})$ is given.
      In the last row, we report the extrapolated values of $\delta$ using Eq.~(\protect\ref{eq:conv}),
      with $x=x_6$.
      }
    \label{tab:app}       % Give a unique label
  \begin{tabular}{l|cc}
    \hline\noalign{\smallskip}
 Basis set & \multicolumn{2}{c}{$p+\tri$ scattering at $E_1=0.20$ MeV} \\
 $n$  & $\delta_n$ (deg) & $x_n$  \\
\noalign{\smallskip}\hline\noalign{\smallskip}
0 & $40.10$ & $-$\\
1 & $42.20$ & $-$ \\
2 & $43.83$ & $0.78$ \\
3 & $45.04$ & $0.74$ \\
4 & $45.98$ & $0.78$ \\
5 & $46.70$ & $0.77$ \\
6 & $47.26$ & $0.78$ \\
\noalign{\smallskip}\hline
Extr. & $49.24$ &  \\
\noalign{\smallskip}\hline
  \end{tabular}
\end{table}

    \end{appendices}

\bmhead{Acknowledgements}    
The Authors would like to acknowledge S. Bacca, N. Barnea, W. Leidemann, and G. Orlandini
for usefull discussions. 


\begin{thebibliography}{99}

\bibitem{A4b}  D.R. Tilley, H.R. Weller, and G.M. Hale,
%               ``Energy levels of light nuclei $A = 4$'',               
               Nucl. Phys. {\bf A541}, 1 (1992)
%
\bibitem{Gatto24} M. Gattobigio and A. Kievsky, Few-Body Syst. {\bf 64}, 86 (2023)
%
\bibitem{E23} E. Epelbaum, Physics {\bf 16} 58 (2023)
%
\bibitem{W70} Th. Walcher,
%              ``Excitation of $\heq$ by inelastic electron scattering at
%                low momentum transfer'',
               Phys. Lett. {\bf B31}, 442 (1970)
%               
\bibitem{Fea65} R.F. Frosch {\it et al.},
%               ``Inelastic electron scattering from $\heq$'',
               Nucl. Phys. {\bf A110}, 657 (1968)
%
\bibitem{Kea83} G. Kobschall {\it et al.},
%             ``Excitation of the quasi-bound state in $\heq$ by electron scattering at
%               medium momentum transfer'',
               Nucl. Phys. {\bf A405}, 648 (1983)
%
\bibitem{Kegel23} S. Kegel {\it  et al.}, Phys. Rev. Lett. {\bf 130}, 152502 (2023)
%               
\bibitem{Hiyama04}  E. Hiyama, B.F. Gibson, and M. Kamimura,
%               ``Four-body calculation of the first excited state of
%               $\heq$ using a realistic NN interaction: $\heq(e,e')\heq(0^+_2)$
%               and the monopole sum rule'',
               Phys. Rev. C {\bf 70}, 031001(R) (2004)
%
\bibitem{Bacca13} S. Bacca, N. Barnea, W. Leidemann, and G. Orlandini,
%                  ``Isoscalar monopole resonance of the alpha
%                     particle: A prism to nuclear Hamiltonians'',
                  Phys. Rev. Lett. {\bf 110}, 042503 (2013)
%
\bibitem{Bacca14} S. Bacca, N. Barnea, W. Leidemann, and G. Orlandini,
%                  ``Is the first excited state of the $\alpha$-particle a
%                  breathing mode?'',
                   Phys. Rev. C {\bf 91}, 024303 (2015)
%
\bibitem{Michel23} N. Michel,W. Nazarewicz and M. Ploszajczak,
  {\tt arXiv:2306.05192}
%
\bibitem{M23} Ulf-G. Meißner, S. Shen, S. Elhatisari, and D. Lee,
%Ab initio calculation of the alpha-particle monopole transition form factor:
  %No puzzle for nuclear forces
             {\tt arXiv:2309.01558}
%
\bibitem{Bogner03} S. Bogner, T. Kuo, and A. Schwenk,
%  Model-independentlow momentum nucleon interaction from phase shift equivalence,
  Physics Reports {\bf 386}, 1 (2003)
%  
\bibitem{AV18} R.B.\ Wiringa, V.G.J.\ Stoks, and R.\ Schiavilla,
               Phys.\ Rev.\ C {\bf 51}, 38  (1995) 
%
\bibitem{EM03} D.R.\ Entem and R.\ Machleidt,
               Phys.\ Rev.\ C {\bf 68}, 041001(R) (2003) 
%
\bibitem{ME11} R. Machleidt and D.R. Entem, Phys.\ Rep.\ {\bf 503}, 1 (2011)
%               
\bibitem{Eea02} E. Epelbaum  {\it et al.},  Phys. Rev. C {\bf 66},
                064001  (2002) 

\bibitem{N07} P. Navr{\'a}til,  Few-Body Syst. {\bf 41}, 117 (2007) 

\bibitem{GP06} A. Gardestig and D.R. Phillips,  Phys. Rev. Lett. {\bf
  96}, 232301 (2006)

\bibitem{GQN09} D. Gazit, S. Quaglioni, and P.  Navr\'atil, 
  Phys. Rev. Lett. {\bf 103}, 102502 (2009)

\bibitem{Mea12} L.E. Marcucci, A. Kievsky, S. Rosati, R. Schiavilla, and M. Viviani,
               Phys.\ Rev.\ Lett. {\bf 108}, 052502 (2012); {\it
                 Erratum}, Phys. Rev. Lett. {\bf 121}, 049901(E) (2018)               

\bibitem{Bea18} A. Baroni  {\it et al.}, Phys. Rev. C {\bf 98}, 044003 (2018)
  

\bibitem{Mea18}  L. E. Marcucci, F. Sammarruca, M. Viviani, and
  R. Machleidt,   Phys. Rev. C {\bf 99}, 034003 (2019)  
%
%\bibitem{Schiavilla} R. Schiavilla, unpublished
%
\bibitem{rep08} A. Kievsky {\it et al.}, J. Phys. G:
   Nucl. Part. Phys. {\bf 35}, 063101  (2008) 
%
\bibitem{fip19} L. E. Marcucci {\it et al.}, {\tt ArXiv:1912.09751}   
%   
\bibitem{Kea01} H. Kamada, A. Nogga, W. Glockle, E. Hiyama, M. Kamimura, K. Varga, {\it et al.},
               Phys. Rev. C {\bf 64}, 044001 (2001) 
%
\bibitem{Viv05} M. Viviani, A. Kievsky, and S. Rosati,
Phys. Rev. C {\bf 71}, 024006  (2005) 
%
\bibitem{bm11} M. Viviani, A. Deltuva, R. Lazauskas, J. Carbonell, A.C. Fonseca, A. Kievsky, L.E. Marcucci, and S. Rosati,
               Phys. Rev. C {\bf 84}, 054010 (2011)
%
\bibitem{bm17} M. Viviani, A. Deltuva, R. Lazauskas, A. C. Fonseca, A. Kievsky, and L. E. Marcucci,
  Phys. Rev. C {\bf 95}, 034003 (2017)
%
\bibitem{Viv20} M. Viviani, L. Girlanda, A. Kievsky, and L.E. Marcucci,
%``$ n+\tri$, $p+\het$, $p+\tri$, and $n+\het$ scattering with the hyperspherical harmonic method'',
  Phys. Rev. C {\bf 102}, 034007 (2020)     
%
\bibitem{zerni} F. Zernike and H.C. Brinkman,
    Proc. Kon. Ned. Acad. Wensch. {\bf 33}, 3 (1935)  
%
\bibitem{F83} M. Fabre de la Ripelle,
    Ann. Phys. (N.Y.) {\bf 147}, 281 (1983) 
%
\bibitem{abra} M. Abramowitz and I. Stegun, {\it Handbook of
    Mathematical Functions} (Dover Publications, Inc., New York, 1970)
%    
\bibitem{Nogga03} A. Nogga, A. Kievsky, H. Kamada, W. Glockle, L.E. Marcucci, S. Rosati, and M. Viviani,
    Phys. Rev. C {\bf 67}, 034004 (2003) 
%
\bibitem{Shen2012} G. Shen, L. E. Marcucci, J. Carlson, S. Gandolfi, and R. Schiavilla
                  Phys. Rev. C {\bf 86}, 035503 (2012)
%
  \bibitem{Pastore13}  S. Pastore, L. Girlanda, R. Schiavilla, and M. Viviani,
    %  "The two-nucleon electromagnetic charge operator in chiral effective field theory ($\chi$EFT) up to one loop",
    Phys. Rev. C {\bf 84}, 024001 (2011)
%
\bibitem{HH08} H.M. Hofmann and G.M. Hale,
%               ``$\heq$ experiments can serve as a database for determining
%               the three-nucleon force'',
               Phys. Rev. C {\bf 77}, 044002 (2008)
    

               %%%%a%%%%%%%%%%  
               
\end{thebibliography}


\end{document}